\documentclass[universe,article,accept,pdftex,moreauthors]{mdpi}

\firstpage{1} 
\pubvolume{1}
\issuenum{1}
\articlenumber{0}
\pubyear{2023}
\copyrightyear{2023}
\externaleditor{Guest Editor: Elisabetta Baracchini}
\datereceived{ } 
\daterevised{ } 
\dateaccepted{ } 
\datepublished{ } 
\hreflink{https://doi.org/} 
\usepackage{silence}
\Title{A Simple Model of Energy Threshold for Snowball Chambers}

\TitleCitation{A Simple Model of Energy Threshold for Snowball Chambers}

\Author{Matthew Szydagis $^{1,}$*\orcidS{}, Cecilia Levy $^{1}$\orcidL{}, Aleksey E.~Bolotnikov $^{2}$, Milind V.~Diwan $^{2}$\orcidD, George J.~Homenides $^{1,\dagger}$\orcidH{}, Alvine C.~Kamaha $^{1,\S}$, Joshua Martin $^{1,\ddagger}$, Richard Rosero $^{2}$, and Minfang Yeh $^{2}$\orcidY{}}

\AuthorCitation{Szydagis, M.; Levy, C.; Bolotnikov A.E.; Diwan, M.V.; Homenides, G.J.; Kamaha, A.C.; Martin, J.; Rosero, R.; Yeh, M.}

\address{%
$^{1}$ \quad Department of Physics, University at Albany SUNY, Albany, NY USA \\
$^{2}$ \quad Brookhaven National Laboratory, Upton, NY USA
}

\corres{Corresponding Author: mszydagis@albany.edu}

\firstnote{Current Address: Department of Physics and Astronomy, University of Alabama, Tuscaloosa, AL USA}
\secondnote{Current Address: Department of Physics and Astronomy, Stony Brook University, Stony Brook, NY USA}
\thirdnote{Current Address: Department of Physics and Astronomy, UCLA, Los Angeles, CA USA}

\abstract{Cloud and bubble chambers have historically been used for particle detection, capitalizing on supersaturation and superheating, respectively. Here we present new results from a prototype snowball chamber, in which an incoming particle triggers crystallization of a purified, supercooled liquid. We demonstrate, for the first time, simulation agreement with our first results from 5 years ago: the higher temperature of the freezing of water and significantly shorter time spent supercooled with respect to control in the presence of a Cf-252 fission neutron source. This is accomplished by combining Geant4 modeling of neutron interactions with the Seitz nucleation model used in superheated bubble chambers, including those seeking dark matter. We explore the possible implications of using this new technology for GeV-scale WIMP searches, especially in terms of spin-dependent proton coupling, and report the first supercooling of WbLS (water-based liquid scintillator).}


\begin{document}

\vspace{-25pt}
\section{Introduction}

The nature of dark matter remains an enduring enigma in cosmology and astroparticle physics. A continued lack of unambiguous evidence from any direct detection experiment of the traditional and well-motivated Weakly Interacting Massive Particle (WIMP) has led to an impetus to consider particle candidates with masses higher and lower than before, driven by many hypothetical models~\cite{cooley2022report}. The main goal of our work is the development~of inexpensive, scalable, supercooled water detectors for the low-mass dark matter search. Auxiliary purposes for particle physics are neutron detection and neutrino studies~\cite{Akimov2017}.

A water target has the advantage of containing hydrogen, ideal for seeking dark matter candidates $O$(1) GeV/c$^2$ in mass due to the recoil kinematics. An additional advantage is the possibility of a high degree of purification, even on large scales~\cite{Fukuda2003}. While having no energy reconstruction, threshold detectors with metastable fluid targets are advantageous for dark matter experiments, due to their high degree of insensitivity to electron recoil~backgrounds (BGs) in the search for nuclear recoil signals, as shown by bubbles chamber experiments such as COUPP~\cite{Behnke2012} and PICO~\cite{Amole2019}. Such detectors rely on particles depositing enough energy $E$ in a critical distance, with the differential energy deposition $dE/dx$ exceeding a critical value. By controlling temperature $T$ and the pressure $P$, the recoil $E$ threshold can remain low while maintaining a high $dE/dx$ threshold.

Supercooled water is a well-motivated medium that has been thoroughly studied~for its own sake~\cite{Dorsey1948,Mossop1955,Holten2014}. Interest in it for particle detection existed years ago~\cite{Agarwal1966,Pisarev1967,Varshneya1969}, and~never for dark matter, until 2016. By contrast, bubble chambers of superheated water are possible, but have too high an $E$ threshold for WIMP detection~\cite{Deitrich1973}. Another motivation for supercooling over superheating is the fact freezing is exothermic not endothermic like boiling; so the phase transition is entropically favorable. The keV-scale recoil $E$ thresholds achieved by dark matter experiments like PICO should be higher than those possible with supercooling, good for low mass. Existing theories suggest even sub-eV threshold is likely at an achievable degree of supercooling~\cite{Barahona2015,Barahona2018,Marcolli2020}, with a high $dE/dx$ threshold for BG e$^{-}$s at the same $T$~\cite{Khvorostyanov2004}.

\section{Materials and Methods}

\begin{figure}[hb]
\includegraphics[width=1\textwidth]{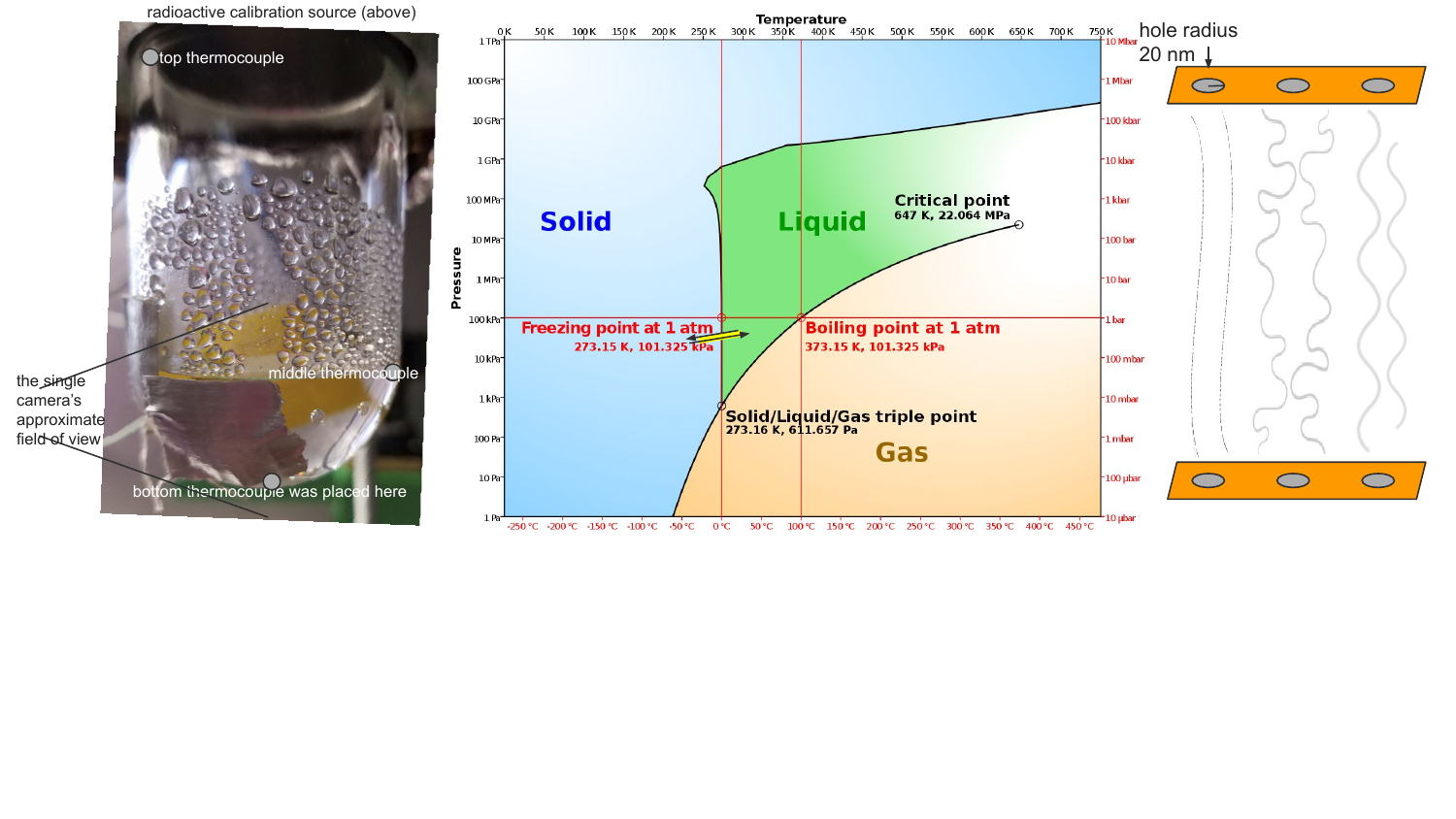}
\vspace{-85pt}
\caption{(\textbf{Left}) The water in the quartz, right after filling (all droplets merged with the volume at bottom after the first cycle). See Fig.~1 of \cite{Szydagis2021} for a greater level of detail. (\textbf{Middle}) The phase diagram of water (courtesy Wikipedia) with our cycle in yellow with arrows. (\textbf{Right}) Simple diagram illustrating the salient features of the ``non-linear'' 20-nm filter: irregular, non-straight holes.}
\label{fig1}
\end{figure}

Water, first deionized to remove metallic impurities, then filtered (150 nm pore size), was boiled. The resulting steam passed through a series of $\mu$m-scale filters before the final one, a 20-nm Novamem PVDF thin-film membrane similar to that used by \cite{Limmer2012}. This last filter remained in place above the water during operation (Figure~\ref{fig1} right). A cylindrical, fused-quartz vessel with hemispherical bottom and top flange for sealing from Technical Glass Products (TGP) was ultrasonicated in an Alconox solution. The ultrasonic cleaning lasted for 15 minutes at 50$^{\circ}$C and 25 kHz. The vessel was then rinsed with the deionized and filtered water and allowed to dry before being sealed up in a Class-1000 cleanroom (Figure~\ref{fig1} left). A low-power vacuum pump reached $\sim$1 psia prior to steam flowing.

The quartz ultimately contained 22 $\pm$ 1 g of purified water and partial vacuum on~top, 8.5 $\pm$ 0.5 psia of water vapor at room $T$. The final mass was limited by the poor throughput of the 20-nm filter, most likely due to particulate build-up. After filling, the quartz vessel was submerged in a Huber / Chemglass ministat circulator for thermal regulation. It was instrumented with three thermocouples for recording vessel $T$ during the cooling process ($T$ ramp) and during the freezing (exothermic-increase $T$s~\cite{Nevzorov2006}). These thermometers were located near the top (below the flange), middle (water line), and bottom (hemisphere tip).

The liquid water was continuously cooled in an ethanol bath to -35$^{\circ}$C at a linear cool-down rate of $\sim$2$^{\circ}$C/min., the best rate for the Huber circulating chiller. While introducing a lag in the water $T$, this had the advantage of reaching a low $T$ and thus higher degree of supercooling rapidly~\cite{Bigg1953}, in an effort to reach low $E$ threshold. The chiller sat on~vibration-dampening pads for prevention of shock-induced nucleation. All data were read in using National Instruments hardware and their LabView software. Data were taken continuously day and night, alternating control (without source) runs versus source runs, to minimize systematics. An effort was made to ensure equal amounts of control and source data were taken. Our cycle is labeled in the phase diagram in the center plot in Figure~\ref{fig1}.

The time spent by the water in the supercooled state  was logged for $^{137}$Cs 662 keV $\gamma$-ray source runs,  $^{252}$Cf neutron and $\gamma$ source runs, and interwoven control runs. The $T$ minima achieved in all cases, prior to exothermic rise, were recorded for all data sets as well. All events were included in the final analysis without non-blind data-quality cuts, to further mitigate bias. The time spent ``active’’ by the water $\Delta t_{active}$ was defined as the time between crossing -15.5$^{\circ}$C and freezing. This value of -15.5 was determined from the $T$ of the severe drop-off in trigger rate in every data set, although analyses using 0$^{\circ}$C as the border instead yielded consistent results in terms of a statistically significant difference between control and Cf. Less time spent supercooled compared to control implies sensitivity to incoming radiation, with freezing at higher $T$ expected for a fixed cool-down rate.

\section{Results}

During a data-taking run of multiple cool-downs for which the $^{252}$Cf source was placed in a reproducible location near the liquid, it did not remain in a metastable \textit{i.e.}~supercooled state as long, freezing also at correspondingly higher $T$. This main result from the seminal snowball chamber was already covered in \cite{Szydagis2021}, which the reader is encouraged to review. In this work, Geant4 (G4)~\cite{Agostinelli2003,Allison2006} Monte Carlo simulations of the nuclear recoil (NR) and electronic recoil (ER) event rates induced by different particle sources placed near the water volume were performed for the purpose of replicating and explaining the data from \cite{Szydagis2021}, which has more detail on the Cs results, which did not differ significantly from control.
\vspace{-5pt}

\subsection{G4 Monte Carlo}

\begin{figure}[H]
\includegraphics[width=1\textwidth]{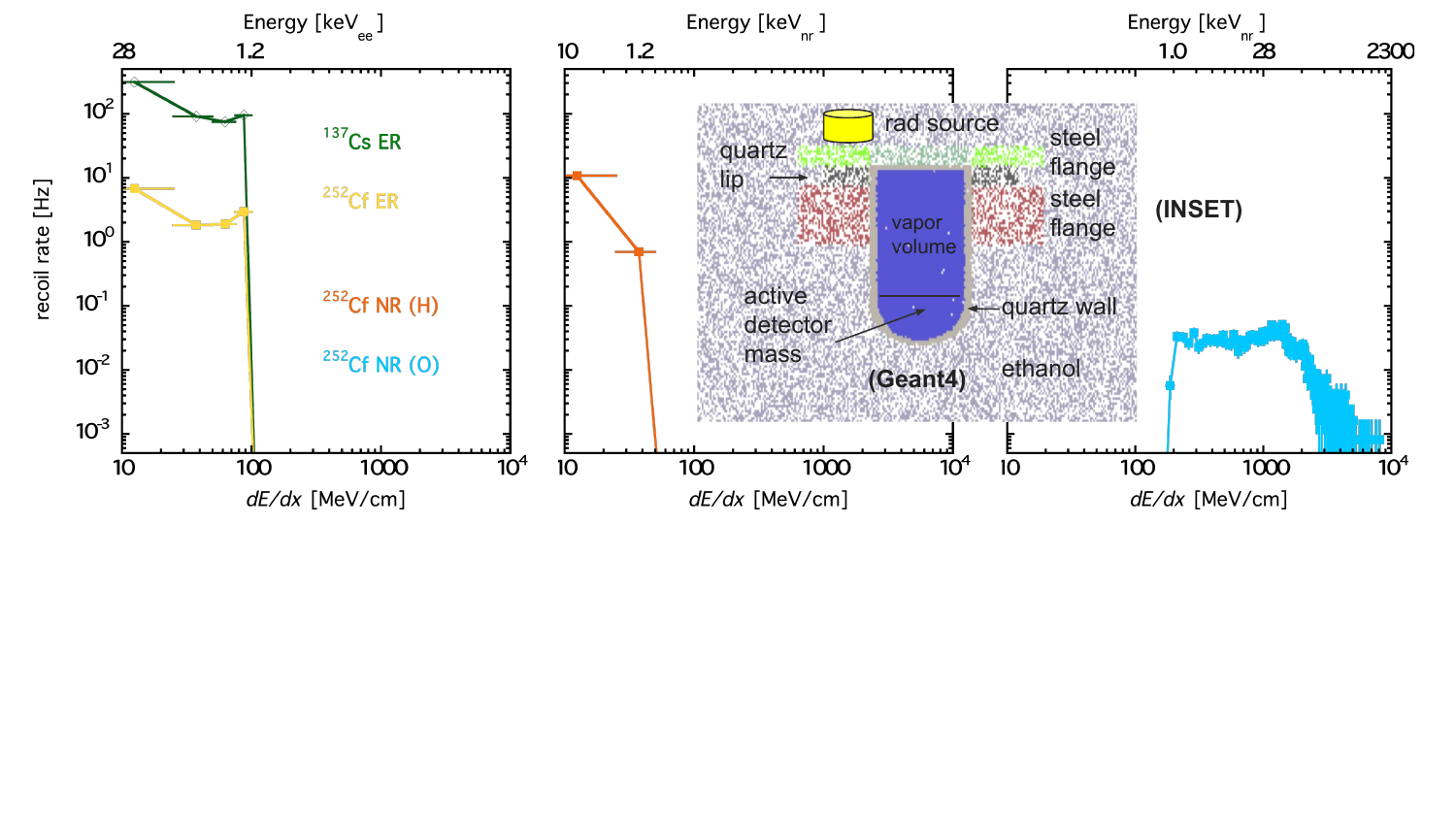}
\vspace{-90pt}
\caption{Stopping power spectra for each possible type of H$_2$O recoil, in the original snowball setup, for tested sources. Corresponding initial species $E$s for which this is the mean $dE/dx$ are along upper x-axes. Plots are, from left to right: e$^-$, p$^+$ (hydrogen), and oxygen recoils, in green or yellow, orange, and cyan, respectively. A $\sim$100 MeV/cm threshold, a natural assumption as explained later in the text, explains a lack of discernible response from a $\gamma$ source (diamonds and line near 100 Hz at upper left in the first pane). Lines are not fits but guides; errors are only bin widths (in $x$) or statistical~($y$). (\textbf{Inset}) G4 geometry: cross-sectional view. Steel flanges blocked $\gamma$s, but ethanol moderated neutrons.}
\label{fig2}
\end{figure}

In this section we summarize the results of G4 Monte Carlo simulations, whereas in the next section we will apply (critical) thresholds in energy, $dE/dx$, and track length on top of these, along with a sigmoid-shaped efficiency for the NR-induced snowball nucleation. Table~\ref{tab1} encompasses all sim results by source; its last two columns are based on Figure~\ref{fig2}.
\vspace{-5pt}

\begin{table}[h] 
\caption{Simulated event rates in the liquid water by source used. For EZAG's $^{252}$Cf, total n rate is based on having a source identical to that in \cite{Dahl2009,Aprile2011}. Source activity at run time is shown along with integrated G4 rates for the active volume, for both sources and recoil types. Parenthetical rates are for $E > 1.2$~keV, $dE/dx > 100$~MeV/cm. Note NR is then strictly oxygen. Upper limits were based on the sim statistics: 1233 and 27.03 seconds of real-time simulated respectively by row. These rates allow us to reproduce the data, in particular those of $^{252}$Cf, after more nucleation conditions are applied.\label{tab1}}
\newcolumntype{C}{>{\centering\arraybackslash}X}
\begin{tabularx}{\textwidth}{CCCCC}
\toprule
\textbf{Calib Source}	& \textbf{Activity [$\mu$Ci]}	& \textbf{Total Rate} & \textbf{NR [Hz] n+($\gamma$,n) } & \textbf{ER Rate [Hz]} \\
\midrule
$^{252}$Cf fission	& 1.0 (all radiation) & $\approx$ 3,000~n/s & 14.2 (2.7) & 13.4 ($<$ 8 x 10$^{-4}$) \\
$^{137}$Cs gammas & 10 (100\% $\gamma$ rays) & 3.7 x 10$^{5}~\gamma$/s & $<$ 4 x 10$^{-2}$ & 570 ($<$ 4 x 10$^{-2}$) \\
\bottomrule
\end{tabularx}
\end{table}
\vspace{-5pt}

As seen at right, the G4 ER rate is 40-50x higher (before/after thresholds) for $^{137}$Cs runs than for $^{252}$Cf, accounting for shielding and geometry (Figure~\ref{fig2} inset and Fig.~2 in \cite{Szydagis2021}). It is unlikely $\gamma$s can explain an enhanced probability for nucleation in $^{252}$Cf's presence.

\vspace{-5pt}

Results from \cite{Varshneya1971} ($^{60}$Co) and ref.~therein can be explained with $T$. These experiments observed $\gamma$s and $\beta$s, but not n, so ER not NR most likely. All thresholds are likely rather sensitive to $T$; these works with data around -20$^{\circ}$C did not quote errors. A small drop in $dE/dx$ threshold (probably only 1-2$^{\circ}$ colder) could steeply increase the Cs ER rate, from negligible to $>$100~Hz. Future work will test higher-$E$ sources, to look for $\gamma$-induced n NR.

This simulated ER vs.~NR discrimination power agrees well with our previous measurements presented in \cite{Szydagis2021}. This power, coupled to low $E$ threshold, would make supercooled water detectors suitable as dark matter experiments. Even if possessing only a keV-scale (not sub-keV) threshold, they would be an improvement over current technologies. A light element, even if only oxygen, does not require as low a threshold as heavier elements~\cite{McCabe2010}. While we have the disadvantage of no visible (yet) proton recoils, upon which was based the idea of sensitivity to sub-GeV WIMPs, one must recall that most ER backgrounds are minimally ionizing ($\approx$2 MeV/cm) so we should be able to decrease the $dE/dx$ threshold safely, lowering $T$. If the snowball chamber acts like a bubble chamber except backwards, colder should imply lower $E$ threshold, as calculated by Barahona~\cite{Barahona2015,Barahona2018} and many others.

\begin{figure}[H]
\includegraphics[width=1\textwidth]{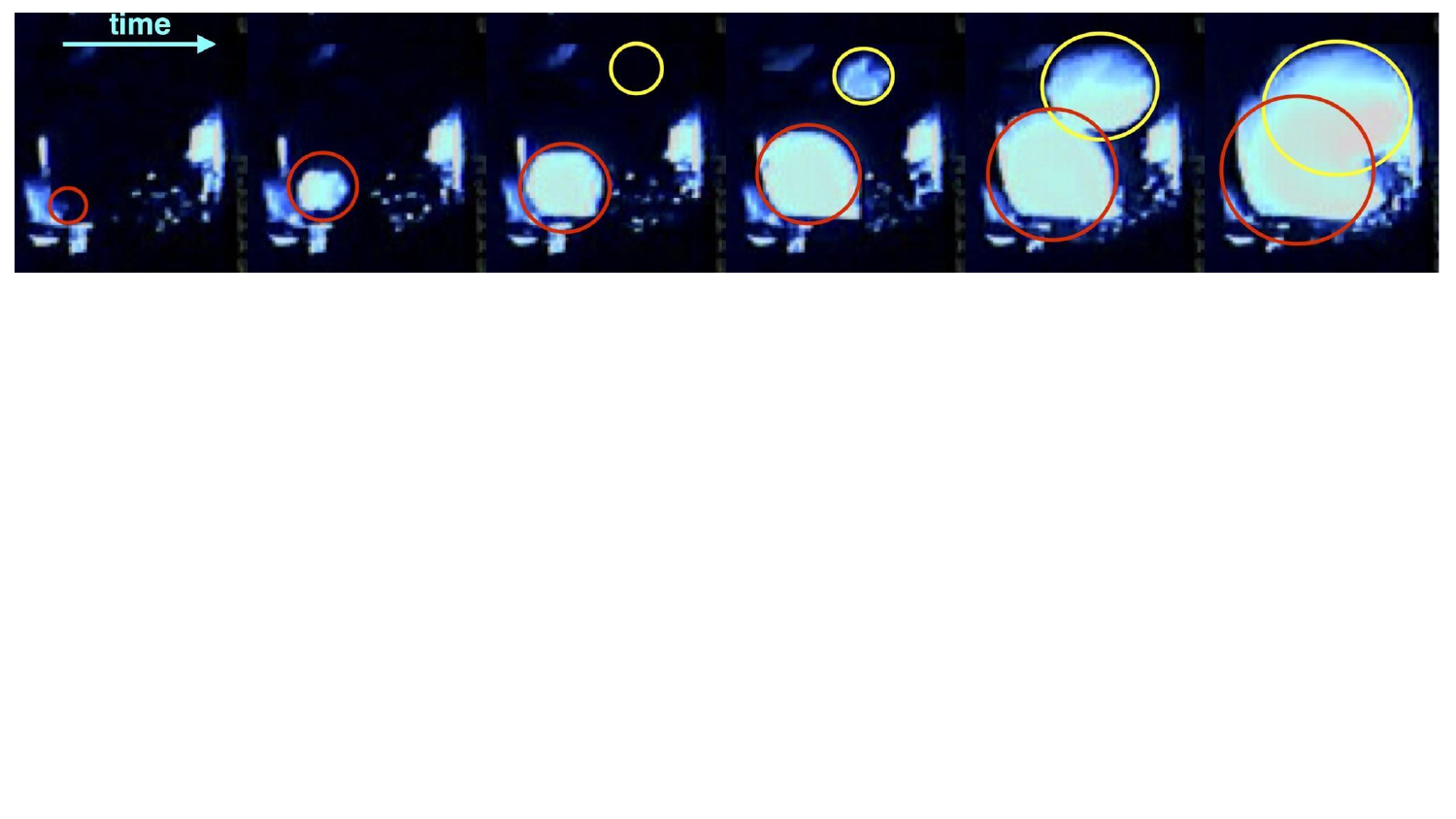}
\vspace{-155pt}
\caption{Images of the crystallization process: 20 mL of supercooled water (primarily black) turning into light-scattering snow (light blue, lit by blue LED) inside a quartz tube. This corresponds to a double-nucleation from neutron data, likely due to multiple scattering. To the best of our knowledge, this was the first time multiple nucleation sites were observed within one sample, in the same image or adjacent frames. Circles mark software detections. Unlike with superheating, nucleations are slow here~\cite{Atkinson2016}. There are 150 ms between the frames depicted. The setup was similar to that of the COUPP 15 kg bubble chamber, with lighting orthogonal to camera line of sight~\cite{Szydagis2010}. Image artifacts included glare, reflections including off the thermocouples on the quartz wall, the wall itself, or the water line, and/or bubbles/dust within the circulating ethyl alcohol surrounding the TGP quartz.}
\label{fig3}
\end{figure}

The results from control and $^{137}$Cs runs were nearly identical~\cite{Szydagis2021}, despite a higher ER rate, suggesting that the snowball chamber, our novel detector, is capable of possessing ER ``blindness'' similar to PICO's~\cite{Amole2019}. The older observations were of needle-like tracks~\cite{Varshneya1969,Varshneya1971}, not spheres as we observed, implying high-energy e$^-$s continuously losing $E$ in the water, and/or multiple scattering from $\gamma$-rays such as a series of Compton interactions and $\delta$-rays. In contrast, all of our results are significantly distinct, individual snowballs which are analogous to the individual bubbles of dark matter bubble chambers and most likely corresponding with neutron elastic scatter vertices (see Figure~\ref{fig3} as well as Figure~5 in \cite{Szydagis2021}).

\subsection{Custom Simulation of Thresholds}

\begin{figure}[ht]
\begin{adjustwidth}{-\extralength}{0cm}
\centering
\includegraphics[width=15.5cm]{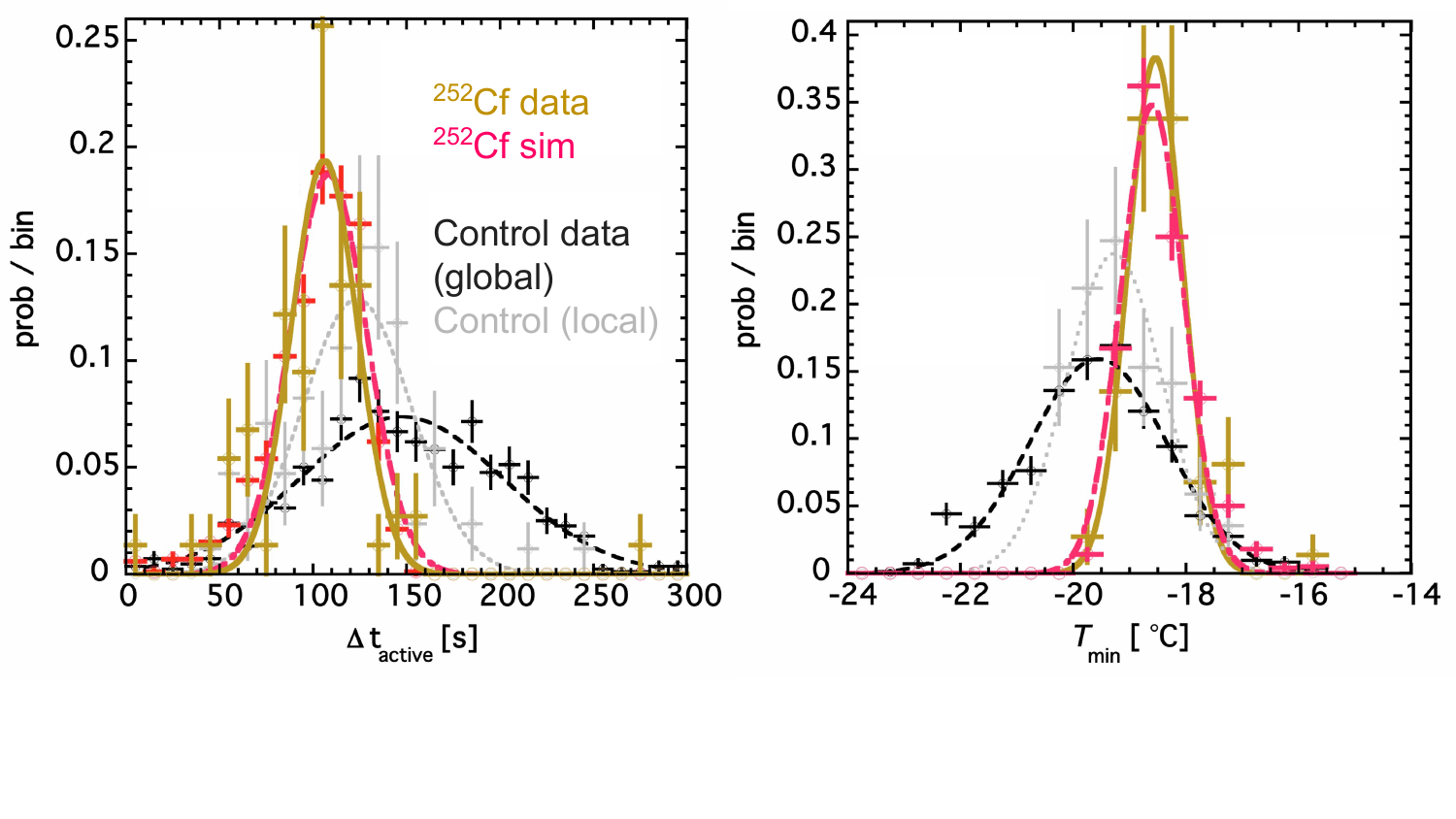}
\end{adjustwidth}
\vspace{-50pt}
\caption{(\textbf{Left}) Histogram of $^{252}$Cf active time (10 s bins) with Gaussian fit, in yellow points and line respectively, compared to control in grey for immediately adjacent runs only. Comparison to them lowers the statistical significance of the disagreement between Cf and control. (That was the comparison used conservatively for our original 5$\sigma$ claim of a discovery of a new physical-chemistry process involving neutrons freezing supercooled water.) However, we also display all control data combined in black, avoiding bias from neutron activation contamination. G4 was used to determine $E$ and $dE/dx$ distributions, with an additional sim adding thresholds and efficiency on top of them: pink. (\textbf{Right}) Same data but looking instead at the $T_{min}$ achieved before crystallization, showing consistent results. 74 Cf and 85 local control events were recorded (840 global control).\label{fig4}}
\end{figure}

It is unclear \textit{a priori} what the best models are for thresholds and the critical radius of nucleation for supercooled fluids. After exploring many similar options, we were able to fit our $^{252}$Cf data using the Seitz ``hot spike'' model that has been used for superheated liquid  bubble chambers for decades~\cite{Seitz1958}. The model agrees well with the data, as seen~in Figure~\ref{fig4}. Figures 2-3 specifically of \cite{Khvorostyanov2004} provided a value for the critical radius of 20~nm, following $S_w$ = 0.97 (a quantity related to the pressure and sample geometry). This radius, held constant during fitting, is also a reasonable assumption because of our filter pore size. A PICASSO-modification-inspired Seitz model~\cite{Behnke2017} (for a superheated~droplet detector or SDD) that best fits our Cf data is codified by these equations, where subscript $c$ is critical:

\begin{flushleft}
(1) \quad $E > E_{c} = 0.2$~keV$_{nr}$ (After conservatively applying Eq.~2-3, $E_{c}$ = 1.2~keV$_{nr}$ effectively) \\
(2) \quad $\frac{dE}{dx} > \frac{E_{c}}{r_{c}} = \frac{200~\mathrm{eV}}{20~\mathrm{nm}}$ = 100~MeV/cm \\
(3) \quad $l > (2r_{c}) = 40$~nm \\
(4) \quad $Efficiency = 1 / [ 1 + ( T / ( 252.8 \pm 1.1~\mathrm{K} ) ) ^ {540~\pm~150} ]$
\end{flushleft}

\noindent
Eq.~(3), ordinarily implied by (2), is necessary~\cite{Szydagis2010}: it ensures a particle traverses at least 1 critical diameter ($l$ is track length, for a 1D approximation). Otherwise, a proto-snowball, like a proto-bubble, may collapse instead of expanding. Eq.~(2) was derived by combining Eq.~(3) with Fig.~4 in \cite{Khvorostyanov2004}. Sub-keV NR has ill-defined $dE/dx$; (2) is assumed to apply to ER.

Applying Eqs.~(2-3) after G4, conservatively to NR too, all ER and NR below $\sim$1~keV are sub-threshold: see Section~3.1, where we presented the resulting $dE/dx$ and $E$ spectra for all of the possible recoiling species: e$^{-}$, H, and O. G4 was cross-checked using NIST for the first two~\cite{ESTAR,PSTAR}, and SRIM~\cite{Ziegler2010} for Oxygen, and found to be consistent. A severe degeneracy in fitting forced us to not float Equations (1-3) above, each already justified by the existing literature, so only a sigmoidal efficiency for nucleation, unrelated to~\cite{Khvorostyanov2004}, was a free (two-parameter) formula. The sigmoid was an expected addition to the Seitz model given~the light elements involved, but it can also be expressed as a softer, sigmoidal energy threshold~\cite{Behnke2013}. Our extreme steepness may explain the reduced nucleation at $T \gtrsim $-16$^{\circ}$C. This could address the fact that (modern) water supercooling researchers are seemingly unaware of the phenomenon of radiation-induced nucleation, known since at least the 1960s-70s if not earlier, as most of the research into supercooled water has taken place at significantly warmer $T$s, far easier to achieve in the lab: less purity is required in terms of critical-radius-sized impurities causing heterogeneous nucleations, in large volumes.

Protons and e$^{-}$s fall below the critical (threshold) $dE/dx$ but not O. This provides a plausible explanation for the lack of decreased supercooled time for Cs, which, despite the thick shielding, like steel flanges, still generated significantly more ER than other sources.

Figure~\ref{fig4} documents a good agreement between Monte Carlo simulation applying a preliminary model and real data on MeV-scale neutrons (which produce keV-scale recoils in H$_2$O) leading to the solidification of water, assuming that it is first supercooled. The data are identical to those in \cite{Szydagis2021}, just binned differently. While not the main DAQ trigger, the top thermometer alone defined the reference $T$s consistently, as in our seminal paper. It was the only thermocouple for which the JB Weld epoxy did not wear off due to repeated thermal cycling in ethanol. The exact $T_{min}$ to use was determined by checking for nucleation in pre-trigger images. (The $T$ increase was the trigger for image recording.) Usage of the top thermocouple was another conservative choice: it was farthest from the water and reacted the weakest/latest (in terms of exothermic rise in $T$). The middle and bottom thermometers exhibited greater statistical significance for the control versus $^{252}$Cf difference.

The control data could not be modeled effectively, so only the sim comparison with $^{252}$Cf is presented. It is not knowable what fraction of control events was due to residual particulate impurities on the order of the critical radius in size and/or wall/surface events (there was no 3D position information, given only one camera), and what fraction due to normal background radiation in the laboratory on the surface, including cosmic muon-induced neutrons. Thus, the control points could only be crudely subtracted. They created another source of systematic uncertainty. As it currently stands, our best fit to the mean time spent active by the water in $^{252}$Cf's presence, 106.1 $\pm$ 2.0 s (data), is 108.3 $\pm$ 0.9 s (sim). The $T_{min}$s agree as well: -18.62 $\pm$ 0.08$^{\circ}$C (data), -18.60 $\pm$ 0.03$^{\circ}$C (sim). (The errors on the simulated quantities are of course smaller, since the statistics can be increased arbitrarily.) Not only the centroids, but the widths are reproduced, and even some of the non-Gaussian skewness, most evident below 50~s in Figure~\ref{fig4} left, although systematics from G4 version and library choices do exist, so we only claim the threshold $E$ is $O$(keV).

Our modified Seitz model was applied continuously as the $T$ ramped down in simulation to mimic real chiller operation, with all threshold conditions dropping with $T$, to follow Figures 2-4 from \cite{Khvorostyanov2004}. Equations (1-3) showed reference values for -20$^{\circ}$C. The fact that Equation (1) is orders of magnitude above even the most pessimistic past calculations of the nucleation activation $E$ suggests a ``quenching factor'' for neutrons may exist, similar to other detector technologies, where NR take more $E$ than ER to produce the same amount of visible signal~\cite{Aprile2011,Aprile2019}. On the other hand, metastable-liquid-based detectors should not exhibit this issue, with NR going directly into the signal channel (the phase transition). Further study is warranted, scanning in $T$ and in neutron kinetic $E$ with mono-energetic neutrons, to better calibrate the threshold or activation $E$.

\subsection{Possible Energy Reconstruction}

\begin{figure}[ht]
\includegraphics[width=0.99\textwidth]{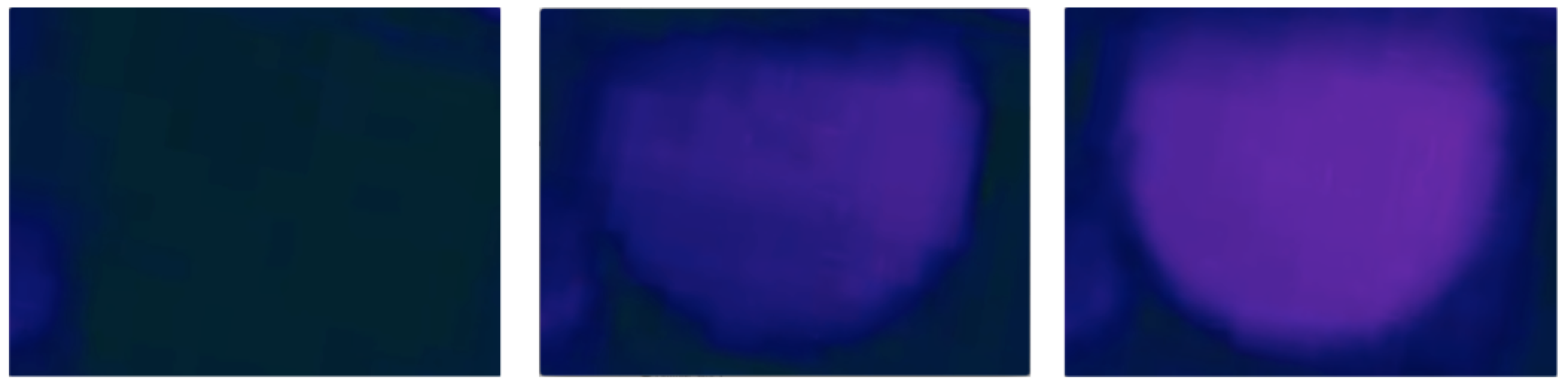}
\vspace{-5pt}
\caption{Thermal imaging capture of supercooled WbLS freezing. These three example frames are ordered chronologically from left to right and are taken from a 25-second-long clip. The color black is approximately -19$^\circ$C, while the purple coloration is 0$^\circ$ on the FLIR unit's internally-calibrated, automatically-adjusting scale. The purple-colored artifact present in all images at left is ice build-up on a thermocouple cable, in the open-air chiller used at BNL. One can see the hemispherical bottom of the test tube come into view very slowly, whereas it was entirely invisible (black) at the start, in thermal equilibrium with the cold air surrounding the vessel. (Note the top oil was already frozen.)}
\label{fig5}
\vspace{-10pt}
\end{figure}

Using a continuous-spectrum source of neutrons like $^{252}$Cf incurs a penalty in terms of systematic uncertainty in the threshold energy and other conditions. But, in addition to future calibrations taking place at a mono-energetic neutron-beam facility (such as TUNL), another possible improvement is the addition of scintillation for direct reconstruction of energy. A scintillating snowball chamber should be analogous to the scintillating bubble chamber, invented by Dahl~\cite{Baxter2017}. Water by itself produces Cerenkov radiation for sufficiently high-$E$ particles~\cite{Fukuda2003} but, unlike noble elements for example such as xenon and argon, it does not scintillate on its own. Water-based liquid scintillator (WbLS), developed by Brookhaven National Laboratory (BNL), addresses this, with scintillation approximately proportional to the amount of $E$ deposited, as expected~\cite{Alonso2014}. An example from the preliminary qualitative results with a snowball chamber based on WbLS instead of water is presented in Figure~\ref{fig5}: in the summer of 2023, a 1~mL sample of WbLS was supercooled at BNL, for the very first time anywhere to the best of our knowledge, without any radioactive sources yet, just to establish the baseline behavior of the fluid. WbLS was successfully cooled to -18.5$\pm0.2^{\circ}$C repeatedly prior to the phase transition occurring.

The micelles of oil-based liquid scintillator inside of the WbLS were too small, being \textit{O}(1 nm), to act as potential nucleators, thus allowing for the deep level of supercooling. No direct visual evidence of the scintillator coming out of solution was observed after 9 cycles, nor degradation in $T_{min}$. In place of vacuum, 1 mL of immiscible, lower-density (mineral) oil on top of the WbLS served as a buffer liquid to maintain water purity and prevent surface nucleation events, in the style of bubble chambers, as rediscovered in the bio-medical community recently, as a means to more deeply supercool water for organ preservation~\cite{Huang2018,deVries2019}. Without oil, the WbLS could only be supercooled (inconsistently) to $\approx$ -10--15$^{\circ}$C, while an identical volume of water of comparable purity in an identical plastic test tube with the same oil buffer froze at -18.9$\pm0.2^{\circ}$C, so the control data point agreed with WbLS on $T_{min}$. Another innovation was the first usage of a FLIR (Forward-Looking Infrared) camera, the Teledyne FLIR One Pro for iPhone, to capture the increase in $T$ of supercooled water during its phase transition. It was operated at 8.5~FPS. Use of a plastic tube in place of quartz allowed for a greater thermal (IR) transmission.

\section{Discussion}
\vspace{-0pt}
This article documents preliminary fits, based on simulations, for data from the 20~g UAlbany snowball chamber that first demonstrated that MeV-scale neutrons (generating keV-scale recoils) can solidify water, after supercooling. Neutrons, x-rays, and other forms of radiation have all been used to study the microscopic properties of supercooled water before in chemistry~\cite{Schiller1971,Hare1990,Belitzky2015} but not its freezing properties for particle physics.

The feasibility of a full-scale water-based dark matter experiment has not yet been established, not until we can further lower the thresholds, so that hydrogen recoils become accessible. However, that should be very easily feasible by going colder through greater purity. Even the lowest $T_{min}$s achieved in the first chamber were still far removed from the coldest possible: $<$ -40$^{\circ}$C for a micro-droplet~\cite{Goy2018}, and -33$^{\circ}$C for a more macroscopic volume~\cite{Hare1987,Hare1990}. Moreover, new filter technology now exists down to 5~nm in pore size.

An increase in the ER background from a reduced threshold in $dE/dx$ could be offset by replacement of water with WbLS, which may not have sufficient light yield to observe low-$E$ NR from WIMPs or neutrons, but does create enough scintillation light (0.1-1~photons/keV, 10-100x below Xe or Ar) to at least act as a veto for high-$E$ background~\cite{Bat2022}, as in \cite{Baxter2017,AlfonsoPita2022}.

$\alpha$ backgrounds such as those from Radon would be mitigated by means of scintillation, or through acoustic discrimination, akin to that discovered by PICASSO then confirmed by COUPP~\cite{Aubin2008,Szydagis2010} (its use continues on PICO), since, unlike most liquids, water expands after it freezes, even given supercooling~\cite{Hare1987}, making it possible that snowballs produce a faint ultrasonic signal, differing by interaction type, in a fashion analagous to expanding bubbles in superheated liquids. $\beta$ and $\gamma$ backgrounds have already been addressed in this article, while $\mu$s can be rejected by Cerenkov light and scintillation, and by means of deep-underground deployment, with active or passive shielding acting as a cosmic-ray muon veto. Neutrons would be guaranteed to multiply scatter in a sufficiently large volume.

Spontaneous nucleation in the bulk can be divided into the categories of heterogeneous and homogeneous. The former can be addressed by robust purification, while the latter~only happens near -40$^{\circ}$C~\cite{Atkinson2016}, if at all (doubt has been cast because past works did not account for particle-induced background nucleation in surface labs). The problem of surface nucleations at the top of the water can mitigated by vacuum, oil, or both. Wall events can be prevented by using radiopure smooth quartz, or hydrophobic vessels or coatings, with fiducialization added. The currently low livetime from melting is improvable by purity, modularity, and deployment deep underground to reduce the cosmic-induced event rate.

\begin{figure}[H]
\includegraphics[width=1\textwidth]{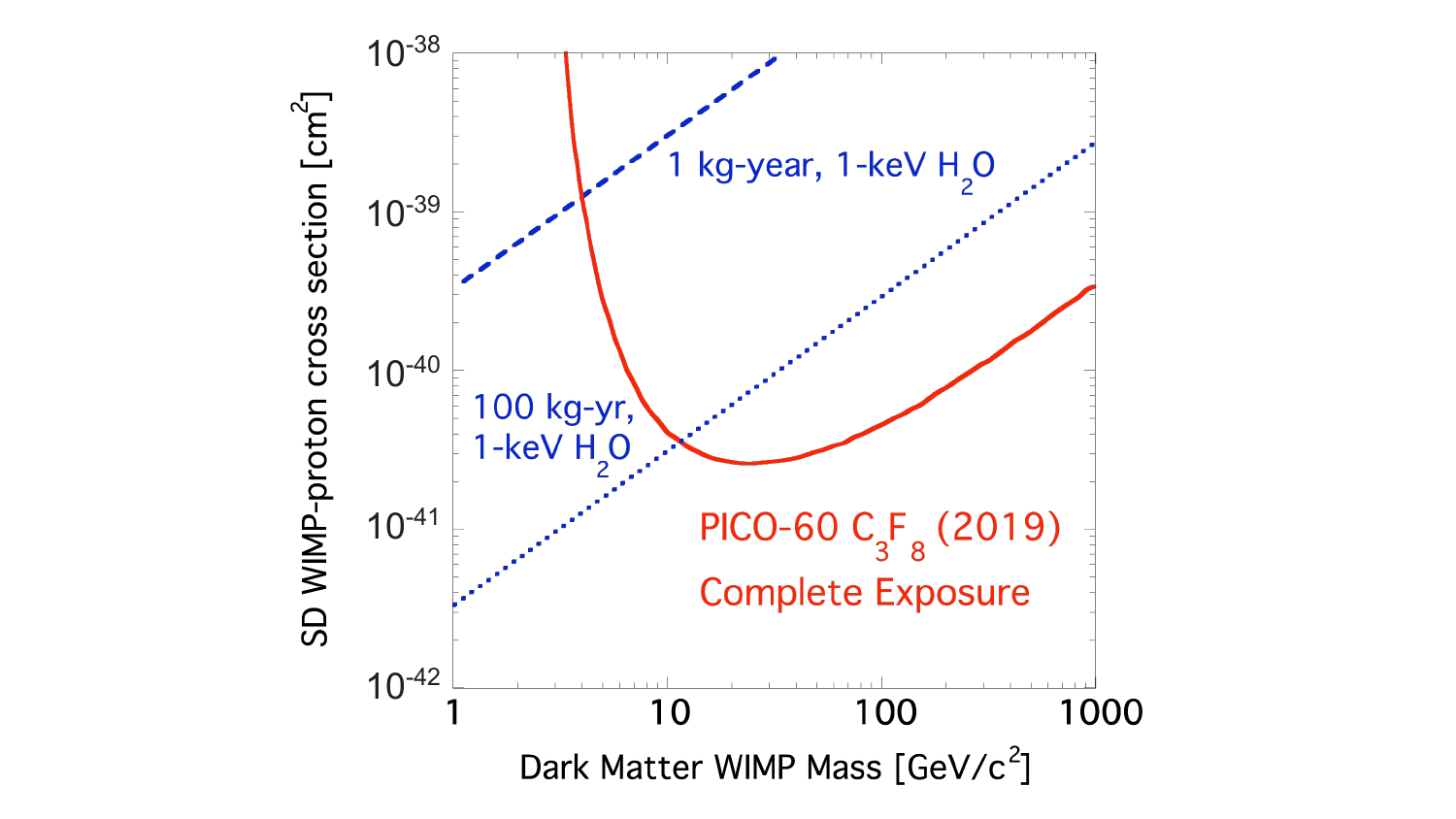}
\vspace{-20pt}
\caption{Projected sensitivities for the spin-dependent interaction of WIMPs with protons as dashed and dotted blue lines for two different snowball-chamber scenarios, compared with the currently world-leading 90\%~C.L.~upper limit on the scattering cross section in this parameter space, the latest result from the PICO bubble chamber experiment~\cite{Amole2019}, depicted as the solid red line. Even 1.0 kg of water underground for only 1 year would probe (SD p) parameter space untouched by any current experiment, projected by a current experiment, or projected for a known planned experiment~\cite{essig2023snowmass2021}. The turn-up due to the $E$ threshold assumed occurs off the scale to the left for this plot, even for 1~keV$_{nr}$, due to H's lightness and its nature (single p$^+$). The same SD model as PICO's was used.}
\label{fig6}
\vspace{-0pt}
\end{figure}

In light of all of these reasonable possibilities for addressing all of the possible background contributions, a strong sensitivity projection is justifiable even if speculative still. While assuming that proton recoil will be measurable, in a background-free environment, we humbly acknowledge our current lack of a precise knowledge regarding the energy threshold, only concluding that its order of magnitude appears to be 1 keV$_{nr}$ based on the only data-sim comparisons currently available. See Figure~\ref{fig6}, still conservative overall.

\section{Conclusions}

We have documented here an initial reproduction of snowball chamber neutron data using the Seitz model for a superheated liquid, with a well-established modification for softening a step-function in efficiency or threshold. MC comparison to the earliest data from $^{252}$Cf is already indicative of an energy threshold of $\sim$0.2-1.2 keV$_{nr}$, at least for oxygen recoils. To best of the authors' knowledge, this was the first attempt in the world, in any sub-field of research related to the supercooling of water, to infer the activation energy for nucleation directly from laboratory data through the introduction of interactions of known energies, as opposed to data on the rate of background events as a function of temperature.

Assuming that supercooling acts like superheating -- except in reverse -- much lower thresholds should be achievable with only slightly lower temperatures than those observed in this work, still a far cry from the lowest known to be possible, even without the modern purification techniques for reducing heteregenous (non-radiation) nucleations.~If possible to achieve without an overwhelming electron-recoil background appearing, sensitivity to recoils on hydrogen nuclei is a valid extrapolation based on this assumption. 

We have also shown that supercooling of water-based liquid scintillator is not only possible, but possible to the same low temperatures as for pure water, another world-first conclusion, which may lead to the creation of the first scintillating snowball chamber.

The combination of properties discussed here are highly suggestive of the utility of~the snowball chamber for direct (WIMP) detection. However, to exceed the present state of~the art in terms of spin-independent WIMP limit-setting, a lower threshold energy must be established first. Fortunately, thresholds as low 1~eV and even lower in energy have been claimed for many years, with multiple different models of existing data on supercooled water~\cite{Khvorostyanov2004,Marcolli2020}. If we conservatively assume that ionization is a necessity and use 12~eV, a snowball chamber would be sensitive to $O(10^{-43})$ cm$^2$ cross-section interactions for the mass range of 0.5-10~GeV/c$^2$. However, as a sub-keV threshold is unproven, it would be premature to show an SI projection. We focus instead on SD proton.

\authorcontributions{Conceptualization, C.L. and M.S; methodology, M.S.; software, M.S.; validation, C.L.; formal analysis, M.S.; investigation, G.J.H.; resources, M.V.D., A.E.B., R.R., and M.Y.; data curation, M.S.; writing---original draft preparation, M.S.; writing---review and editing, A.C.K.; visualization, J.M.; supervision, M.S. and C.L.; project administration, M.S.; funding acquisition, C.L. All authors have read and agreed to the published version of the manuscript.}
\vspace{-1pt}
\funding{This research was funded by the U.S.~DOE (grant\# DE-SC0015535) and by a Project SAGES Award to Prof.~Levy. The work conducted at Brookhaven National Laboratory was supported by the DOE under contract DE-AC02-98CH10886. Prof.~Szydagis thanks BNL Instrumentation Division director Dr.~Gabriella Carini for hosting him.}
\vspace{-1pt}
\dataavailability{The data presented in this study are available upon request.} 
\vspace{-1pt}
\acknowledgments{The authors thank Dr.~Peter Wilson for his assistance in the publication of the first snowball paper. We also acknowledge Dr.~Ernst Rehm and Claudio Ugalde of Argonne National Laboratory for useful discussions regarding their water bubble chamber for nuclear astrophysics. We thank Prof.~Kathy Dunn of CNSE (College of Nanoscale Science \& Engineering) for performing electron microscopy on the filtering membrane.}
\vspace{-1pt}
\conflictsofinterest{The authors declare no conflict of interest.}
\vspace{-1pt}

\begin{adjustwidth}{-\extralength}{0cm}

\reftitle{References}

\bibliography{refSnow.bib}
\PublishersNote{}
\end{adjustwidth} \end{document}